\documentclass[fleqn,usenatbib,useAMS]{mnras}

\usepackage{newtxtext,newtxmath}

\usepackage[T1]{fontenc}
\usepackage{ae,aecompl}

\usepackage{graphicx}	
\usepackage{amsmath}	
\usepackage{amssymb}	
\usepackage[section]{placeins}

\newcommand{\der}{{\rm d}}
\defcitealias{GC19}{GC19}

\title[The Endgame of Gas Giant Formation]{The Endgame of Gas Giant Formation: \\Accretion Luminosity and Contraction Post-Runaway}

\author[S. Ginzburg and E. Chiang]{
Sivan Ginzburg$^{1}$\thanks{E-mail: ginzburg@berkeley.edu}\thanks{51 Pegasi b Fellow.}
and Eugene Chiang$^{1,2}$
\\
$^{1}$Department of Astronomy, University of California at Berkeley, CA 94720-3411, USA\\
$^{2}$Department of Earth and Planetary Science, University of California at Berkeley, CA 94720-4767, USA
}

\date{Accepted XXX. Received YYY; in original form ZZZ}

\pubyear{2019}

\begin{document}
\label{firstpage}
\pagerange{\pageref{firstpage}--\pageref{lastpage}}
\maketitle

\begin{abstract}
Giant planets are thought to form by runaway gas accretion onto solid cores. Growth must eventually stop running away, ostensibly because planets open gaps (annular cavities) in their surrounding discs. Typical models stop runaway by artificially capping the accretion rate and lowering it to zero over an arbitrarily short time-scale. In reality, post-runaway accretion persists as long as the disc remains. During this final and possibly longest phase of formation, when the planet is still emerging from the disc, its mass can more than double, and its radius contracts by orders of magnitude. By drawing from the theory of how gaps clear, we find that post-runaway accretion luminosities diverge depending on disc viscosity: luminosities fall in low-viscosity discs but continue to rise past runaway in high-viscosity discs. 
This divergence amounts to a factor of $10^2$ by the time the disc disperses. Irrespective of the specifics of how planets interact with discs, the observed luminosity and age of an accreting planet can be used to calculate its instantaneous mass, radius, and accretion rate. We perform this exercise for the planet candidates embedded within the discs orbiting PDS 70, HD 163296, and MWC 758, inferring masses of 1--10 $M_{\rm J}$, accretion rates of 0.1--10 $M_{\rm J}$/Myr, and radii of 1--10 $R_{\rm J}$. Our radii are computed self-consistently from the planet's concurrent contraction and accretion and do not necessarily equal the value of $2R_{\rm J}$ commonly assumed; in particular, the radius depends on the envelope opacity as $R \propto \kappa^{0.5}$.
\end{abstract}




\section{Introduction}\label{sec:introuction}

The leading theory for the formation of giant planets is the core accretion model, in which a rocky or icy core several times the mass of Earth gravitationally accretes a gas atmosphere from an ambient circumstellar disc \citep{PerriCameron1974,Harris1978,Mizuno1978,Mizuno1980,Stevenson1982,BodenheimerPollack86,Pollack96}.
Initially, the bottleneck for gas accretion is the atmosphere's Kelvin--Helmholtz (KH) cooling time, i.e. how long atmospheric gas takes to radiate away its gravitational energy, thereby contracting onto the core and allowing fresh nebular gas to take its place 
\citep[e.g.][]{PisoYoudin2014}. This cooling time, which sets the atmosphere's mass-doubling rate, increases as the atmosphere grows \citep[e.g.][]{LeeChiang2015}. Once the atmosphere and core become comparable in mass and atmospheric self-gravity can no longer be neglected, the KH growth time decreases with increasing mass: the planet accumulates mass over ever shorter time-scales in a `runaway' growth phase \citep[][and references therein]{BodenheimerPollack86,Pollack96,Ikoma2000,Lee2014,PisoYoudin2014,LeeChiang2015,Piso2015,GC19}. During 
pre-runaway and runaway, the 
radius of a nascent gas giant is several orders of
magnitude larger than the radius of Jupiter $R_{\rm J}$,
as the planet's atmosphere extends to the 
Bondi radius (or the Hill radius if it is smaller) 
where it connects to the external nebula 
\citep{Bodenheimer2000}.

At some stage, the nebula is unable to supply gas at a sufficient rate to keep pace with the continuously shortening
cooling time-scale. Accretion then becomes controlled hydrodynamically
rather than thermodynamically.
\citet[][hereafter \citetalias{GC19}]{GC19} considered the hydrodynamics
of gas flows in the sub-thermal regime,
appropriate for planets whose Bondi radii
are less than their Hill radii (which in turn
are less than the disc pressure scale height).
For sub-thermal planets---these could be to up
to several times the mass of Jupiter $M_{\rm J}$
in the outermost
portions of discs, beyond 10 au---the hydrodynamic
accretion rate should be given by Bondi 
\citep[e.g.][]{Edgar2004}.
At fixed ambient disc density, growth at the Bondi rate
still runs away, as the mass-doubling time-scale
continues to decrease with increasing mass.

What stops runaway may be the opening of gaps:
annular depletions of disc density around the planet's
orbit effected by repulsive Lindblad torques 
\citep{GoldreichTremaine80,GoodmanRafikov2001,GinzburgSari2018}.
The more massive the planet, the deeper the gap it
excavates and the more it staves off its mass supply \citep{LinPapaloizou93,Bryden99,Kley99,Lubow99,DAngelo03,TanigawaIkoma2007,Lissauer2009,Machida2010,TanigawaTanaka16,Lee2019};
this negative feedback loop lengthens the mass-doubling time-scale until it exceeds the gas disc lifetime  
of a few million years  \citep{Mamajek2009,WilliamsCieza2011,Alexander2014}.
Throttling of gas accretion is most
severe in low-viscosity discs where gaps are
deepest \citepalias[][and references therein]{GC19}.

Our goal here is to calculate how the radius $R$ and 
luminosity $L$ of a planet evolve during this last 
phase of nebular accretion, after the planet
transitions out of runaway. How does a planet
shrink from its Bondi 
radius, and
what is its peak luminosity, still powered by residual accretion? 
Our calculation applies directly to
planets just emerging from
and still feeding off their parent discs,
of which there are
now several candidates directly
imaged \citep{Sallum2015,Guidi2018,Reggiani2018,Wagner2018,Haffert2019}. We will estimate the masses and accretion rates of these objects.

While our analysis is directly  
inspired by a modern understanding of
how gaps open in discs, we keep much of our treatment
general by using a simple power-law parametrization
of post-runaway
accretion that can be adjusted to model
different disc-planet interaction scenarios.
Regardless of the details of any particular scenario,
post-runaway accretion 
should play out naturally over the lifetime
of the gas disc---and so we will find
planet contraction and luminosity histories
unfolding over the same long time-scale of Myrs
(by contrast to earlier, artificially short 
accretion prescriptions made for computational
convenience; see section 2.1 of \citealt{Marley2007}).
Most of the planet's contraction, amounting to
orders-of-magnitude reduction in radius,
occurs during this final stage of nebular accretion.

The rest of this paper is organized as follows. 
In Section \ref{sec:cooling} we 
review how cooling-limited accretion runs away.
The subsequent hydrodynamically-limited phase, when
accretion continues through gaps, is discussed 
in Section \ref{sec:hydro}. There we describe
our method for calculating the post-runaway
radius, luminosity, and effective temperature.
Section \ref{sec:tracks} presents the resultant
evolutionary tracks, both as functions of
mass and of time, with dependences on opacity
noted in Section \ref{sec:opacity}.
Applications to observations are made in Section 
\ref{sec:observations},
after which we summarize
and provide an outlook in Section \ref{sec:summary}.

\section{Runaway cooling}\label{sec:cooling}

The cooling-limited phase of gas accretion onto planetary cores has been studied extensively, with an emphasis on pre-runaway accretion, 
when
the planet's gas atmosphere is lighter than the core 
\citep{Pollack96,Ikoma2000,PapaloizouNelson2005,Rafikov2006,Lee2014,PisoYoudin2014,LeeChiang2015,Piso2015,Ginzburg2016,Lee2018}. 
In \citetalias{GC19} we extended the analytical scaling relations of 
pre-runaway cooling
to the runaway regime, when the atmosphere surpasses the core in mass and dominates the gravitational field. 
Here we repeat and expand upon the calculation of runaway cooling.
We omit
order-unity 
coefficients
to concentrate on scaling relations.

During runaway, most of the nascent giant's mass $M$ resides in its self-gravitating convective atmosphere (an ideal-gas
polytrope with an adiabatic index $\gamma$).
The convective interior is overlaid by 
a radiative layer whose temperature varies by a factor
of order unity from
the surrounding nebula's temperature $T_0$.
We define the planet's radius $R$ as the location of the 
radiative--convective boundary (rcb).
The planet's accretion rate is initially given by the KH cooling time-scale $t_{\rm cool}\equiv E/L$. The gravitational energy 
to be radiated away 
is $E\sim GM^2/R$ where $G$ is the gravitational constant,
and the luminosity 
follows from 
applying the diffusion equation to the radiative layer: $L\sim 4\upi R^2\sigma T_0^4/\tau$,
where $\sigma$ is the Stefan--Boltzmann constant and $\tau$ is the optical depth at the rcb 
measured radially from the outside in \citep[for power-law opacities, $T^4$ changes by an order-unity factor across the radiative layer; e.g.][]{Rafikov2006,PisoYoudin2014,GinzburgSari2015}.

From hydrostatic equilibrium, the temperature at the planet's centre is given by $k_{\rm B}T_{\rm c}\sim GM\mu/R$, where $k_{\rm B}$ is Boltzmann's constant and $\mu$ is the mean  molecular weight.
The central density of the polytrope is given by $\rho_{\rm c}\sim M/R^3$. We use the adiabatic relation $\rho\propto T^{1/(\gamma-1)}$ to calculate the density at the rcb: 
\begin{equation}\label{eq:rho_rcb}
\rho_{\rm rcb}=\frac{M}{R^3}\left(\frac{k_{\rm B}T_0R}{GM\mu}\right)^{1/(\gamma-1)}=\frac{M}{R^3}\left(\frac{R}{R_{\rm B}}\right)^{1/(\gamma-1)},
\end{equation}
where
\begin{equation}
R_{\rm B}=\frac{GM\mu}{k_{\rm B}T_0}
\end{equation}
is the Bondi radius, and where we substituted $T_0$ for the temperature at the rcb. 
The optical depth at the rcb is
\begin{equation}\label{eq:tau}
\tau=\kappa\rho_{\rm rcb}h=\frac{\kappa M}{R R_{\rm B}}\left(\frac{R}{R_{\rm B}}\right)^{1/(\gamma-1)},
\end{equation}
where $\kappa$
and $h=R^2/R_{\rm B}$ are the local
opacity and scale height, respectively.
The cooling time-scale is therefore
\begin{equation}\label{eq:t_cool}
t_{\rm cool}=\frac{E}{L}=\left(\frac{k_{\rm B}}{\mu}\right)^5\frac{\kappa}{4\upi\sigma G^4}\frac{T_0}{M^2}\left(\frac{R_{\rm B}}{R}\right)^{(4\gamma-5)/(\gamma-1)}.
\end{equation}

We adopt a nominal disc temperature of $T_0=100\textrm{ K}$ for consistency with \citetalias{GC19}. As explained below, our main results are independent of this choice.
We assume for simplicity a nominal constant rcb dust opacity $\kappa = 0.1\textrm{ cm}^2\textrm{ g}^{-1}$,
neglecting weak dependences on temperature and density in the relevant parameter range
\citep{BellLin94,Piso2015};\footnote{Specifically, we are motivated by the nominal opacities of \citet{Piso2015}, as given by the dashed red line in their fig. 16; these account for grain growth following \citet{DAlessio2001}.} we also test a lower dust-free opacity
in Section \ref{sec:opacity}.
We take $\mu=2\textrm{ amu}$, appropriate for molecular hydrogen at the rcb. 
As long as KH cooling is the bottleneck for gas accretion, $R \sim R_{\rm B}$ to within a logarithmic
factor \citep{PisoYoudin2014,Ginzburg2016}; in other words, the rcb scales with the atmosphere's
outer boundary, which is located at $R_{\rm B}$ for the wide-orbit planets of interest here \citepalias[see][]{GC19}.
Hence, during this stage the planet's radius increases as $R\propto M$ and its luminosity as $L\propto M^3$. The growth time decreases as $t_{\rm cool}\propto M^{-2}$ according to
equation \eqref{eq:t_cool}, indicating runaway gas accretion. 

We note that as long as $R\sim R_{\rm B}$, the planet's interior temperature is of order $T_0$, implying that $\gamma=7/5$ (hydrogen is molecular) throughout the planet. As the planet contracts and heats up in later stages, $\gamma$ varies spatially, necessitating a numerical integration (Section \ref{sec:numerical}).

\section{Hydrodynamic Regulation}\label{sec:hydro}

While KH cooling is one limit to the planet's growth rate, another is the nebula's ability to supply the planet with fresh gas. We parametrize the hydrodynamical mass-doubling time-scale with a power law
$t_{\rm hydro}\equiv M/\dot{M}\propto M^\beta$;
this is equivalent to a parametrization of the accretion rate $\dot{M}$ with time. 
In \citetalias{GC19} we found that Bondi accretion through a planet-carved gap corresponds to $\beta=3$ in viscous discs (\citealt{ShakuraSunyaev73} viscosity parameters $\alpha\gtrsim 10^{-3}$) 
and $\beta\approx 15$ in nearly inviscid ones ($\alpha\lesssim 10^{-3}$). These results were obtained by calculating the time history of gap depletion by the planet's repulsive gravitational torque. In Section \ref{sec:tracks} we examine $\beta=3$ and $\beta=15$ as fiducial cases, and demonstrate that the two values lead to qualitatively different outcomes. In Section \ref{sec:observations} we treat $\dot{M}$ as a free
function
to explore other limiting mechanisms, 
also hydrodynamical but not necessarily involving gaps \citep{Szulagyi2014,TanigawaTanaka16}. 

Because $t_{\rm cool} \propto M^{-2}$ (during runaway) and $t_{\rm hydro} \propto M^\beta$ where $\beta$ is
generally positive, there is a transition point when the time-scales cross and accretion becomes limited hydrodynamically
rather than thermodynamically \citepalias[see fig.~1 of][]{GC19}. We denote by $M_0$
the planet mass at this transition: 
\begin{equation}\label{eq:t_hydro}
t_{\rm hydro}\equiv \frac{M}{\dot{M}}=t_{\rm cool}(M_0)\left(\frac{M}{M_0}\right)^\beta \,.
\end{equation}
The transition marks the end of runaway and the starting point for the calculations
in the remainder of this paper. 
For simplicity, unless otherwise
indicated, we adopt a fixed
$M_0 = 0.5 M_{\rm J}$, characteristic
of values computed in \citetalias{GC19}.\footnote{Equating equations (9) and (18) of \citetalias{GC19}, and assuming their disc model, we find $M_0\propto M_{\rm disc}^{0.2}\kappa^{0.2}\alpha^{0.28}a^{0.09}$ for $\beta=3$ and $M_0\propto M_{\rm disc}^{0.29}\kappa^{0.06}a^{0.64}$ for $\beta=15$; $a$ is the planet's orbital radius and $M_{\rm disc}$ is the total mass of the disc, which determines its density normalization.} 
The parametrization in equation \eqref{eq:t_hydro}
enables us to study the sensitivity of our results to both $\beta$ and the minimum growth time $t_{\rm cool}(M_0)$ (evaluated using equation \ref{eq:t_cool} with $R=R_{\rm B}$).

We now show that after the transition, post-runaway, the
planet adjusts so that $t_{\rm cool} = t_{\rm hydro}$.
Suppose this were not true, that just after the transition 
$t_{\rm cool} < t_{\rm hydro}$.
For times $t_{\rm cool} < t < t_{\rm hydro}$, 
the planet radiates away its gravitational energy and undergoes KH contraction while initially keeping a constant mass
(since growth is throttled on the longer time-scale $t_{\rm hydro}$).
The planet thus detaches from the nebula as $R<R_{\rm B}$ \citep[see also][]{Bodenheimer2000,Marley2007}.  
Now according to equation \eqref{eq:t_cool}, as the planet contracts, 
$t_{\rm cool}\propto R^{-(4\gamma-5)/(\gamma-1)} \propto R^{-3/2}$ increases. Eventually $t_{\rm cool}$ reaches $t_{\rm hydro}$ (whose value does not depend on $R$),
whereupon the planet resumes its growth.
We conclude that post-runaway, the planet grows and contracts simultaneously (see Fig.~\ref{fig:mass}) while
satisfying the condition $t_{\rm cool}=t_{\rm hydro}$.

\subsection{Numerical scheme for $R(M)$}\label{sec:numerical}

The planet's contraction during its post-runaway growth $R(M)$ can be calculated by solving
$t_{\rm cool}(M,R)=t_{\rm hydro} (M)$. This
equation has to be solved numerically, since $\gamma$ is not
uniform as hydrogen dissociates and ionizes during the planet's contraction and heating. We now describe our general solution 
scheme, and defer to 
the appendix
analytical solutions for some asymptotic cases.

The planet's post-runaway evolution is divided 
into two successive stages.  During the first stage, the planet remains partitioned by an rcb (Section \ref{sec:radiative}), while in the second stage it is fully convective (Section \ref{sec:convective}). 
These stages correspond to the `stalling' and `cooling' regimes in section 4.2 of \citet{Berardo2017}.

\subsubsection{Post-runaway stage 1: radiative envelope}\label{sec:radiative}

We denote by $T$ the temperature at the planet's photosphere. Initially, $T$ equals the nebular temperature $T_0$, and the planet is engulfed by a thick, radiative, and nearly isothermal envelope of optical depth $\tau\gg 1$, similar to
conditions during runaway (Section \ref{sec:cooling}).
Now as then, the central temperature and density are given by $k_{\rm B}T_c=GM\mu/R$ and $\rho_c=M/R^3$.
We take the temperature at the rcb to be identical to the photospheric
temperature $T$, and find the rcb density 
$\rho_{\rm rcb}$ by numerically integrating 
\begin{equation}\label{eq:deriv}
\frac{\der \ln\rho}{\der \ln T'}=\frac{1}{\gamma(\rho,T')-1}
\end{equation}
from $(T_{\rm c},\rho_{\rm c})$ to $(T,\rho_{\rm rcb})$ using a fourth-order Runge--Kutta 
method. The adiabatic index $\gamma$ is given by
equations (B2) and (B3) of \citet{Piso2015}, 
supplemented by the Saha equation for calculating
the fractions of molecular, atomic, and ionized hydrogen.
In Section \ref{sec:tracks} we detail how
hydrogen
transitions from molecular ($\gamma=7/5)$ to atomic ($\gamma=5/3)$ within the planet's evolving envelope, with $\gamma$
dropping in regions of partial
dissociation or ionization as energy goes into
breaking molecules or atoms instead of heat \citep{Saumon95,Lee2014,Piso2015}. Having solved for $\rho_{\rm rcb}$, we estimate
the optical depth at the rcb as 
\begin{equation} \label{eq:rhokappah}
\tau(M,R,T)=\rho_{\rm rcb}\kappa h
\end{equation}
where $h=(k_{\rm B}TR^2)/(GM\mu)$ is the scale height there.

The planet cools with luminosity
\begin{equation}\label{eq:lum_radiative}
L=\frac{4\upi R^2\sigma T_0^4}{\tau}=\frac{GM^2}{t_{\rm hydro} R},    
\end{equation}
where the second equality follows from $t_{\rm cool}=t_{\rm hydro}$ (see the 
paragraph just before Section \ref{sec:numerical}).
It follows that the optical depth must also satisfy
\begin{equation}\label{eq:tau_hydro}
\tau(M,R,T_0)=\frac{4\upi R^3\sigma T_0^4t_{\rm hydro}(M)}{GM^2} \,.
\end{equation}
We solve equation \eqref{eq:tau_hydro} numerically
for $R(M)$, calculating the left-hand side
using equation \eqref{eq:rhokappah} independently of the right-hand side.

We note in passing that 
for the radius to shrink from $R_{\rm B}$ to much smaller values (smaller by two orders of magnitude according to Fig. \ref{fig:mass}), the planet 
almost certainly has to find a way to shed its spin
angular momentum.
Possible mechanisms for angular momentum loss include
expulsion of material into a circumplanetary disc \citep{WardCanup2010} and magnetic interaction between the planet and such a disc \citep{TakataStevenson96,Batygin2018}. We
do not model these processes and merely assume
that they 
are efficient enough that the planet contracts on a KH time-scale. 

\subsubsection{Post-runaway stage 2: fully convective}\label{sec:convective}

As the planet shrinks, 
it must radiate away an accretion flux
$L/(4\upi R^2)\propto M^2/(t_{\rm hydro} R^3)$
given by (\ref{eq:lum_radiative})
that increases; see the appendix, in particular equations
\eqref{eq:rad1} and \eqref{eq:tau_m}, for 
why the factor of $R^3$ dominates.
To meet this need, the optical depth at the rcb
decreases (so that $L/(4\upi R^2) = \sigma T_0^4/\tau$
increases). 
When $\tau=1$, the planet is no longer able to radiate away the accretional energy with a radiative layer at nearly the nebular temperature $T_0$. The planet is now fully convective with a rising photospheric temperature $T>T_0$ given by 
\begin{equation}\label{eq:temp}
L=4\upi R^2\sigma T^4=\frac{GM^2}{t_{\rm hydro} R}\,.
\end{equation} 

There has been a longstanding 
debate
in the literature as to 
whether the temperature at the planet's 
photosphere
is correctly given by equation \eqref{eq:temp}
(see \citealt{Berardo2017} for a review).
Since the planet radius lies inside the Bondi radius,
the infalling gas is supersonic and
has its bulk kinetic energy converted into heat through an accretion shock. The equilibrium temperature behind the shock depends on 
how efficiently the post-shock gas cools,
and can range 
from $T_0$ (isothermal with the nebula; this would
be the case if all the post-shock gas
lost its energy to photons that escaped
without thermalizing) to $T$ 
as
given by equation \eqref{eq:temp} (complete
thermal equilibrium between radiation and matter
in the post-shock region). For more details,
see section 2.1 of \citet{Commercon2011} 
and also \citet{Zeldovich}.
These two 
limits 
lead to different 
specific entropies, and by extension
different cooling luminosities after planets stop accreting
(`cold starts' vs. `hot starts'; \citealt{Fortney2005,Fortney2008,Marley2007,SpiegelBurrows2012,Berardo2017}). 
Deciding between the two limits
during the radiative-envelope stage (Section
\ref{sec:radiative}) 
is not important
because the accretion luminosity $L<4\upi R^2\sigma T_0^4$ as given by equation \eqref{eq:lum_radiative} cannot 
lift the photospheric temperature $T$ substantially above the nebular floor $T_0$. For the subsequent fully convective stage, we assume the hot limit as given by equation \eqref{eq:temp} holds. 
Our calculation is thus
compatible with classical high-entropy hot starts 
\citep[such models also seem to better fit directly imaged planets, although this might be a result of selection bias;][]{Bowler2016,SnellenBrown2018,Wang2018,Dupuy2019,Nielsen2019}. 
Specifically, in our model $t_{\rm cool}$ equals $t_{\rm hydro}$, 
and both
equal $t_{\rm disc}$ when the planet reaches its final mass. The planet's formation entropy is therefore the same as that of a classical hot start evaluated at an age of $t_{\rm disc}$ \citep[hot-start models cool down from arbitrarily hot initial conditions
which are
eventually
forgotten; see e.g.][]{Fortney2008}.
The detailed radiation-hydrodynamics
calculations of \citet{Marleau2017} and \citet{Marleau2019} 
rule out cold starts and yield temperatures within an order-unity factor of our equation \eqref{eq:temp}.

The evolution of $R(M)$ during the convective-envelope stage is found numerically by asserting the fully-convective condition: at the rcb,
\begin{equation}\label{eq:tau_condition_hot}
\tau(M,R,T)=1    
\end{equation}
with $T$ satisfying $t_{\rm cool}=t_{\rm hydro}$ according to equation \eqref{eq:temp}, and with the left-hand side calculated using 
equation \eqref{eq:rhokappah}
(but now with $T>T_0$).

\section{Evolutionary Tracks}\label{sec:tracks}

\begin{figure}
	\includegraphics[width=\columnwidth]{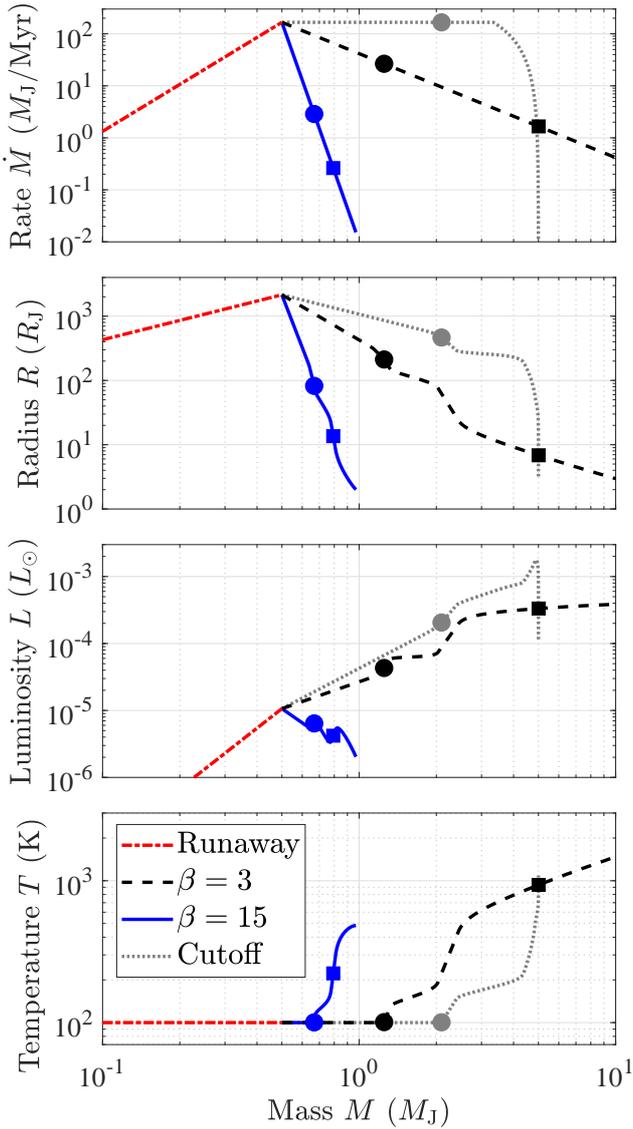}
	\caption{The accretion rate, radius, luminosity, and photospheric temperature of a planet 
	as it grows in mass. 
	Initially, for $M < M_0 = 0.5 M_{\rm J}$, the planet extends to the Bondi radius and accretes mass at a runaway pace set by the Kelvin--Helmholtz time $t_{\rm cool}$, evaluated using an rcb opacity $\kappa= 0.1\textrm{ cm}^2\textrm{ g}^{-1}$ (dot--dashed red lines; see Section \ref{sec:cooling}). For $M>M_0$, accretion is limited by the hydrodynamical time $t_{\rm hydro}\equiv M/\dot{M}\propto M^{\beta}$,
	where $\beta=3$ pertains to high disc viscosities and $\beta=15$ to low ones \citepalias{GC19}. During this stage the planet simultaneously
	accretes mass and contracts, satisfying $t_{\rm cool}(M,R)=t_{\rm hydro}(M)$, an equation we solve in Section \ref{sec:numerical}. Vertical kinks in the curves correspond to hydrogen dissociation and ionization in the planet's interior.
	Filled circles mark the transition to a fully-convective envelope, after which $T$ equals the thermalized accretion value given by equation \eqref{eq:temp} and
	rises above the nebular $T_0=100\textrm{ K}$.
	Filled squares mark the planet's final mass, when $t_{\rm hydro}$ exceeds the protoplanetary gas disc's lifetime $t_{\rm disc}=3\textrm{ Myr}$.
	The dotted grey line depicts an additional model in which the accretion rate is kept constant and then artificially cut off to reach a final mass of $5 M_{\rm J}$.}
	\label{fig:mass}
\end{figure}

\begin{figure} 
	\includegraphics[width=\columnwidth]{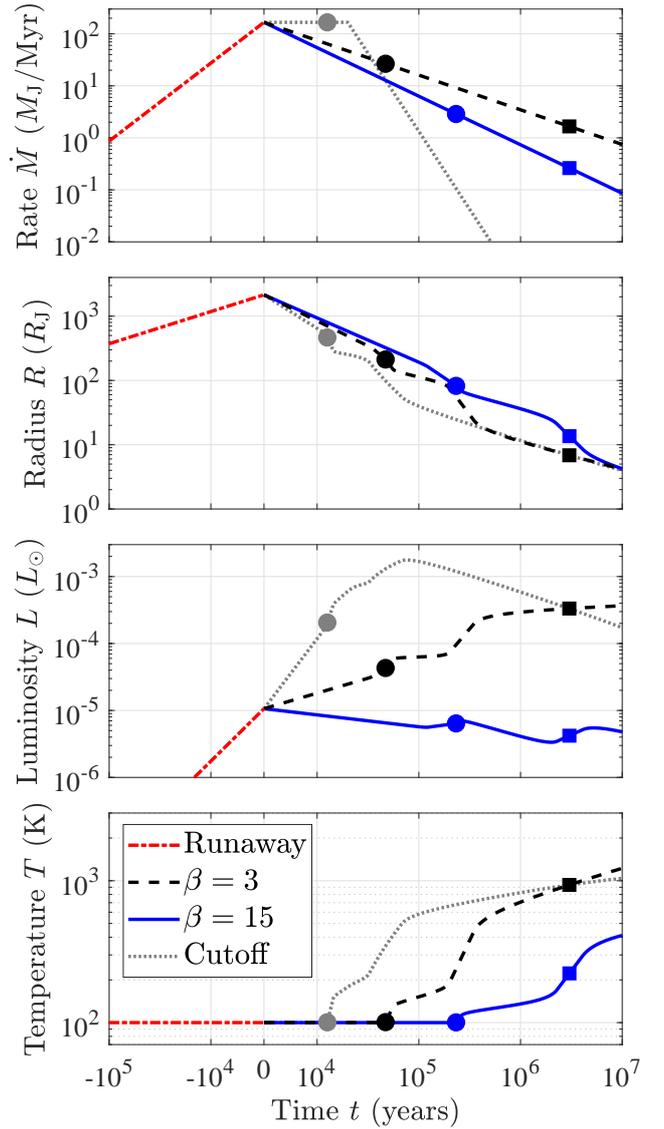}
	\caption{Same as Fig. \ref{fig:mass}, but as a function of time.
	The transition from runaway cooling to
	gap-mediated hydrodynamic accretion is marked by $t=0$;
	at this moment, $M = M_0 = 0.5 M_{\rm J}$, $R = R_{\rm B}$, and $t_{\rm cool}=t_{\rm hydro}\approx 3\times 10^3\textrm{ yrs}$ (equation \ref{eq:t_cool} with $\kappa= 0.1\textrm{ cm}^2\textrm{ g}^{-1}$). The time axis is logarithmic for both $t>0$ and $t<0$,
	but with times $|t|<t_{\rm cool}(M_0)$ near the transition
	not plotted.
	During runaway cooling, $-t \propto M^{-2}$ for $M \ll M_0$
	(Section \ref{sec:cooling}).
	During the hydrodynamic phase, $+t \propto M^\beta$ (Section \ref{sec:hydro}). The accretion rate of the `cutoff' model is set to decay as $\dot{M}\propto t^{-3}$. During the decay $M/\dot M\gg t$, and the contraction luminosity surpasses the accretion luminosity.}
	\label{fig:time}
\end{figure}

\begin{figure}
	\includegraphics[width=\columnwidth]{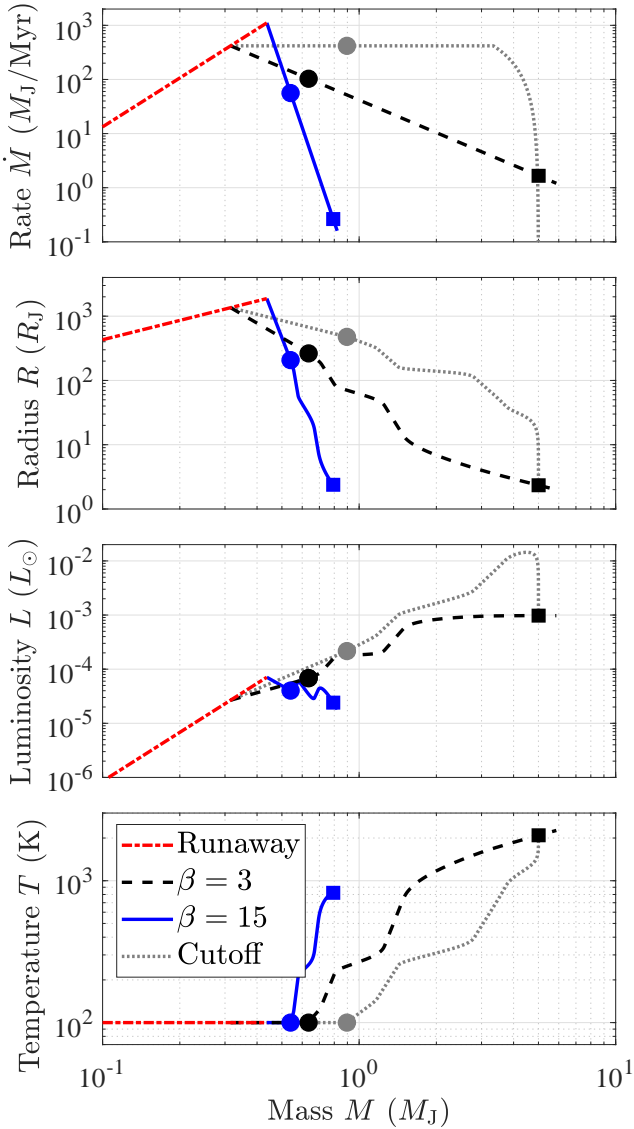}
	\caption{Same as Fig. \ref{fig:mass}, but with $\kappa=10^{-2}\textrm{ cm}^2\textrm{ g}^{-1}$
	appropriate for a dust-free rcb. 
	We keep the same final masses as in Fig. \ref{fig:mass} by changing $M_0$: $M_0(\beta=3)\approx 0.32M_{\rm J}$ and $M_0(\beta=15)\approx 0.44M_{\rm J}$.}
	\label{fig:kappa}
\end{figure}

\begin{figure}
	\includegraphics[width=\columnwidth]{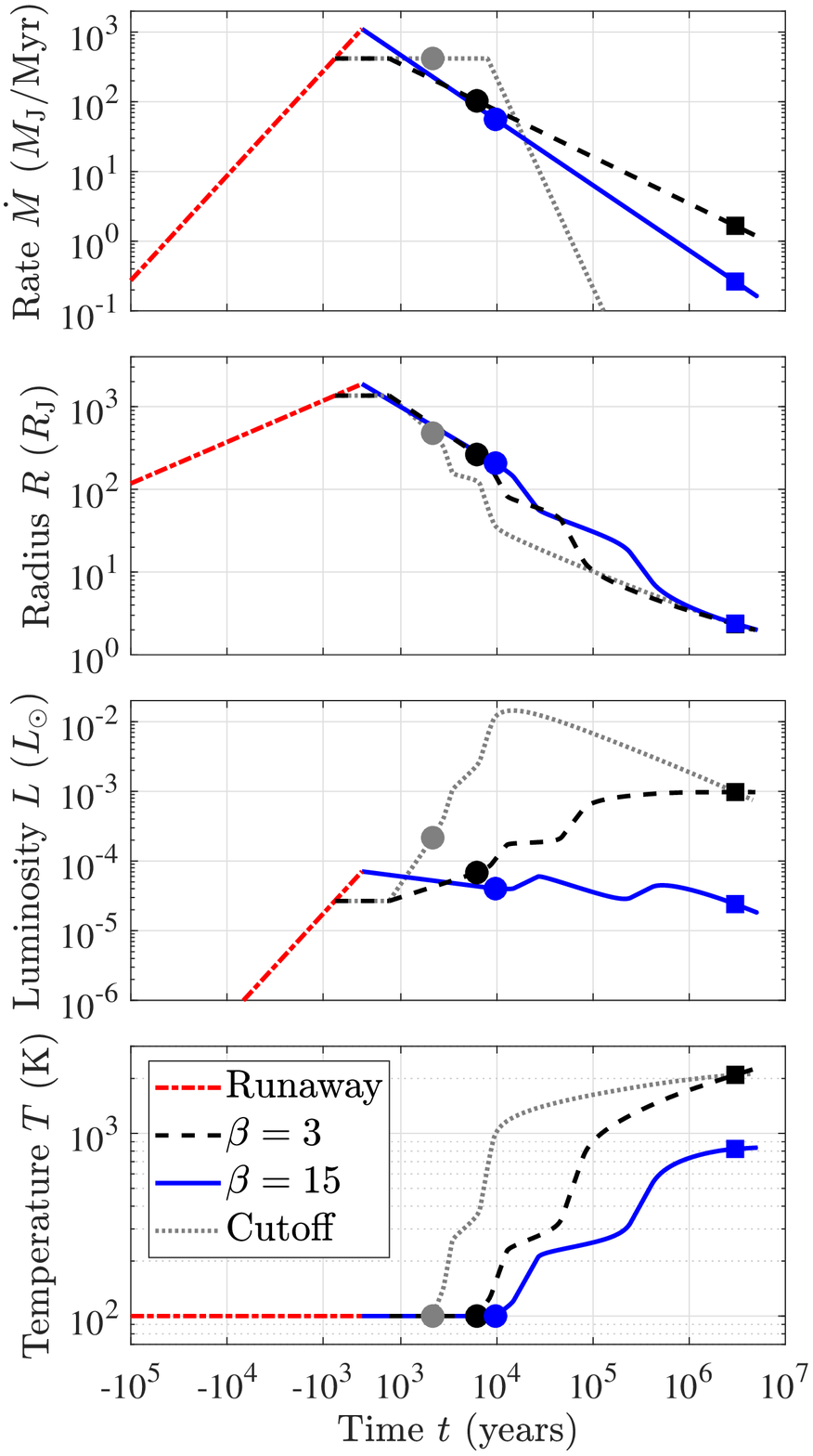}
	\caption{Same as Fig. \ref{fig:time}, plotted with the same
	logarithmic time axis excised for times near $t=0$, but computed using $\kappa=10^{-2}\textrm{ cm}^2\textrm{ g}^{-1}$ (so this figure is also the same as Fig.~\ref{fig:kappa}, plotted here as a function of time). 
	Compared to $\beta=15$, the minimal KH cooling time-scale for the $\beta=3$ model is now longer and so its 
	transition to the hydrodynamic phase is slower; 
	hence the 
	flat segment of the dashed curve that extends just past $t=0$.}
	\label{fig:kappa_time}
	
\end{figure}

Figs \ref{fig:mass} and \ref{fig:time} 
show how our nominal planet's 
radius, luminosity, and photospheric temperature 
evolve with mass and with time.
At a mass of $M_0=0.5\, M_{\rm J}$, the planet transitions from cooling-limited runaway growth at $R = R_{\rm B}$ (dot--dashed red lines; Section \ref{sec:cooling}) to hydrodynamically-limited 
mass growth on a time-scale $t_{\rm hydro} \propto M^{\beta}$ 
and concurrent contraction to $R < R_{\rm B}$
(Section \ref{sec:hydro}).
Results are plotted for two values of $\beta$ describing accretion in gaps opened in high viscosity discs ($\beta = 3$) and low viscosity discs ($\beta = 15$) \citepalias[see][]{GC19}. 

As the planet contracts, 
its interior heats up ($k_{\rm  B}T_c\sim GM\mu/R$),
causing 
first molecular hydrogen to dissociate and next 
atomic hydrogen to ionize.
These two phase transitions lower the adiabatic index
$\gamma$ and thereby accelerate contraction
(see equations \ref{eq:rad1} and \ref{eq:rad_hot});
the photospheric temperature and luminosity 
jump
during these transitions.\footnote{Self-gravitating atmospheres 
become dynamically unstable when $\gamma<4/3$ in a large enough portion of the planet.
Therefore, the phase transitions might occur on even shorter time-scales than are shown in Fig. \ref{fig:time} (namely dynamical time-scales).}
As a guide to interpreting some of the behaviour
exhibited in Fig. \ref{fig:mass}, we provide in the appendix an analytical derivation of the asymptotic radius and luminosity evolution in two regimes: prior to dissociation, and after the interior is fully ionized. 

The filled circles in Figs \ref{fig:mass} and \ref{fig:time} mark the transition to fully convective envelopes 
whose photospheric temperatures exceed the nebular temperature $T_0$ (see bottom panels, and Section \ref{sec:convective}). From
this point onward ($\sim$$10^5$ years after the end of runaway),
the evolution 
does not depend on our choice of $T_0$.
The filled squares indicate the planet's final mass, attained when $t_{\rm hydro} = M/\dot{M}$ equals the nominal gas disc lifetime $t_{\rm disc}=3\textrm{ Myr}$ 
(we extend the curves in Fig.~\ref{fig:time} to consider $t_{\rm disc}$ up to $10^7$ years).
The filled squares mark 
where our calculation ends and where cooling models for
isolated (non-accreting) planets begin.
The case $\beta=3$ naturally leads to a higher final mass.
In Fig.~\ref{fig:mass}, we truncate the curves
at $R = 2 R_{\rm J}$, the approximate radius for which 
electron degeneracy and Coulomb interactions, neglected in our
equation of state, become important.
Since the planet reaches its final mass at $R \approx 10 R_{\rm J}$,
it is safely modelled as an ideal gas while accreting
(and for some time after accretion ceases).

Although the accretion rate peaks at the end of runaway
(see discussion surrounding equation \ref{eq:t_hydro} and the top panels of Figs \ref{fig:mass} and \ref{fig:time}),
the luminosity does not necessarily peak there;
$L$ can either decrease or increase
after the planet transitions from cooling-limited
to hydro-limited accretion,
depending on the value of $\beta$.
This result can be understood analytically (see 
equations \ref{eq:beta_cr_cold} and \ref{eq:lum_hot}).
A few Myrs after the end of runaway, when accreting planets are most
likely to be observed, the $\beta = 3$ and $\beta = 15$ models
are separated by two orders of magnitude in luminosity: $L\approx 4\times 10^{-4}L_\odot$ and $L\approx 4\times 10^{-6}L_\odot$, respectively, where $L_\odot$ is the solar luminosity.
These luminosities are roughly constant while
the planet accretes from $1\textrm{ Myr} \lesssim t\lesssim 10\textrm{ Myr}$. 

Our luminosity evolution lacks the distinctive bright $\gtrsim 10^{-3}L_\odot$ flash found in many earlier computations \citep{Bodenheimer2000,Hubickyj2005,Marley2007,Mordasini2012}.
This brief flash is an artefact of the ad hoc way that runaway growth is terminated,
as pointed out by \citet{Bodenheimer2000}. In these models, the prescribed accretion rate is kept 
near its maximal value during most of the planet's contraction and then switched off rapidly 
so as to arrive at a final mass near Jupiter \citep[see e.g. fig. 2 of][]{Mordasini2012}.
The sharp rise in luminosity ($L \sim G M \dot{M}/R$) 
occurs because $R$ is allowed to drop while $\dot{M}$ is
still pinned by hand to its maximum value.
By contrast, in our models, $R$ and $\dot{M}$ decrease simultaneously, resulting in a flatter luminosity curve.
The accretion rate in \citet{Mordasini2017} also drops roughly as a power law from its maximal value, yielding a luminosity curve that lacks the short bright flash 
(their fig. 1). For further illustration, we add a third model (dotted grey line) to Figs \ref{fig:mass} and \ref{fig:time}. This model terminates the accretion in a similar way to previous studies, demonstrating that the bright flash is an artefact. We choose to cut off this model at the same final mass as our $\beta=3$ model to emphasize that the radius and luminosity of a planet at a given age and mass do not depend on its accretion history---they are determined solely by KH cooling. 
After such an unphysically sharp cutoff, $M/\dot{M}\gg t$ and the planet cools and contracts essentially as if it were not accreting at all.

\section{Dependence on opacity}\label{sec:opacity}

From Figs \ref{fig:mass} and \ref{fig:time},
the photospheric temperature (= rcb temperature) 
can approach 
the dust sublimation point.
Grain growth and sedimentation
in planetary envelopes can also
decrease the dust opacity  \citep{Movshovitz2010,Mordasini2014,Ormel2014}. 
Furthermore, the low opacities of dust-free gas may 
explain how the exceptionally voluminous
atmospheres of low-mass `super-puff' 
planets accreted (\citealt{LeeChiang2016}, their section 4).
For these reasons we are motivated to repeat our
computation with a lower rcb opacity.

Figs \ref{fig:kappa} and \ref{fig:kappa_time}
are the same as Figs \ref{fig:mass} and \ref{fig:time} but
calculated using
$\kappa=10^{-2}\textrm{ cm}^2\textrm{ g}^{-1}$ 
which roughly characterizes dust-free gas
for our 
temperatures and pressures
\citep{Freedman2008}. 
Cooling faster with a lower opacity, 
planets contract to approximately 2--3 $R_{\rm J}$ during the last few Myrs of their formation. This result reproduces previous calculations that employed similar opacities \cite[e.g.][]{Mordasini2012} and can be understood using equation \eqref{eq:kappa_r}: for a given mass and age, the radius scales approximately as $R\propto\kappa^{8/17}$.
Below about $2 R_{\rm J}$,
our neglect of degeneracy pressure 
and Coulomb interactions (`cold terms') 
in the equation of state renders our calculation
less reliable 
(Jupiter-mass planets have a minimum `cold' radius of about $R_{\rm J}$).

Giant planets 
newly emerging 
from their parent discs are commonly assumed to have 
$R \approx 2 R_{\rm J}$ \citep[e.g.][]{Eisner2015,Zhu2015}.
Our analysis shows that this assumption
is justified only for dust-free atmospheres. In Section
\ref{sec:observations} we 
further explore the impact of $\kappa$ 
on the interpretation of observations.

\section{Interpreting Observations}\label{sec:observations}

There are currently several candidates for planets that are still accreting gas from their parent  circumstellar discs (\citealt{Sallum2015,Guidi2018,Reggiani2018,Wagner2018,Haffert2019};
cf.~\citealt{Mendigutia2018,Currie2019}). The
observed
luminosity $L$
of a given candidate 
is usually translated into a constraint on $M\dot{M}$ (the planet's mass times its accretion rate) using $L=GM\dot{M}/R$, where the planet's radius $R$ is either assumed to be 1--2 $R_{\rm J}$ \citep{Eisner2015,Zhu2015}, or left as a free parameter. 
Planetary radii during
post-runaway accretion are indicated in
Figs \ref{fig:time}
and \ref{fig:kappa_time}; in particular,
for high envelope opacities appropriate to dusty
gas, 
planets a few Myrs old could still
be several times larger than Jupiter. 

In Section \ref{sec:hydro} we explained that the
planet cools and contracts on the same time-scale that it grows in mass:
$t_{\rm cool}(M,R)=t_{\rm hydro}\equiv M/\dot{M}$.
From this we
computed $R(M,\dot{M})$ and $L(M,\dot{M})$ for specific accretion rates $\dot{M}(M)$ (equivalently $\dot{M}(t)$); the latter
were physically 
motivated by 
how planets accrete from within disc gaps. 
While 
accretion luminosities
can be factors of $\sim$$10^2$
higher in viscous discs than in
inviscid ones,
we caution that observing a high luminosity 
does not necessarily rule out
inviscid discs,
as all the curves in Figs \ref{fig:mass} and \ref{fig:time} 
shift as a function of
the 
parameter $M_0$,
the mass for which the planet
transitions from cooling-limited
to hydrodynamically-limited
accretion. 
To illustrate this degeneracy,
increasing $M_0$ from our nominal value of $0.5 M_{\rm J}$
to $3 M_{\rm J}$ in Fig. \ref{fig:mass}
would extend the runaway line and increase 
post-runaway
luminosities by a factor of $10^2$ 
(see also the last paragraph of the appendix).
That said, the observed planet candidate
orbiting MWC 758 is so bright ($L>10^{-3} L_\odot$;
\citealt{Reggiani2018}) 
that we cannot reconcile it with our $\beta=15$ and $\kappa=0.1\textrm{ cm}^2\textrm{ g}^{-1}$
inviscid track (but the $\beta =3$ viscous
track is compatible).
Given the candidate's distance from its host star
of $\sim$20 au,
we estimate using the disc model of \citetalias{GC19} 
that $M_0\approx 1M_{\rm J}$ if $\beta = 15$
(see also the footnote in Section \ref{sec:hydro}),
which implies that its peak luminosity does not exceed $10^{-4}L_\odot$. 
We will give another reason why the inviscid
scenario is not compatible
with MWC 758 when we infer its mass below.

\begin{figure}
	\includegraphics[width=\columnwidth]{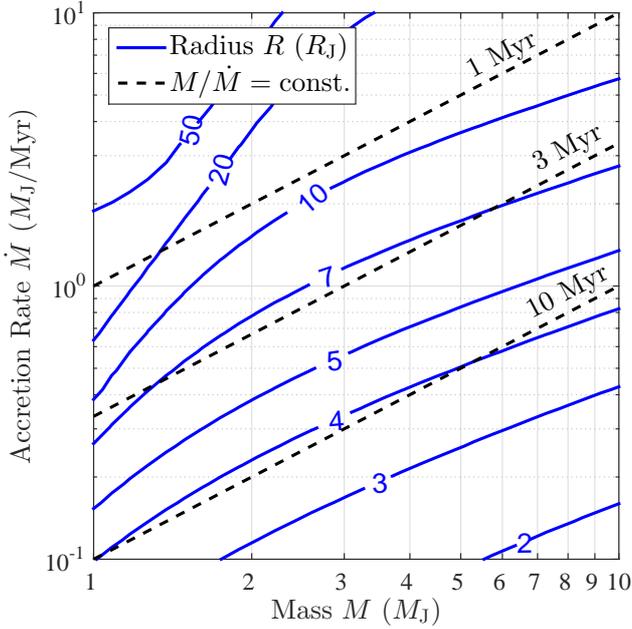}
	\caption{Contours of constant radius $R(M,\dot{M})$ (solid blue lines) for an
	accreting protoplanet 
	with a dusty rcb opacity $\kappa= 0.1\textrm{ cm}^2\textrm{ g}^{-1}$. 
Radii are computed by equating the planet's Kelvin--Helmholtz cooling time to its 
	growth time: $t_{\rm cool}(M,R)=t_{\rm hydro}\equiv M/\dot{M}$. 
This is the same equation used to compute previous figures,
	but it is used here
	with an arbitrary $\dot{M}$ (i.e. no relation $\dot{M}(M)$ and in particular no
	$\beta$ are assumed).  
	Dashed black lines indicate mass doubling times $M/\dot{M}$.}
	\label{fig:rad_contour}
\end{figure}

\begin{figure}
	\includegraphics[width=\columnwidth]{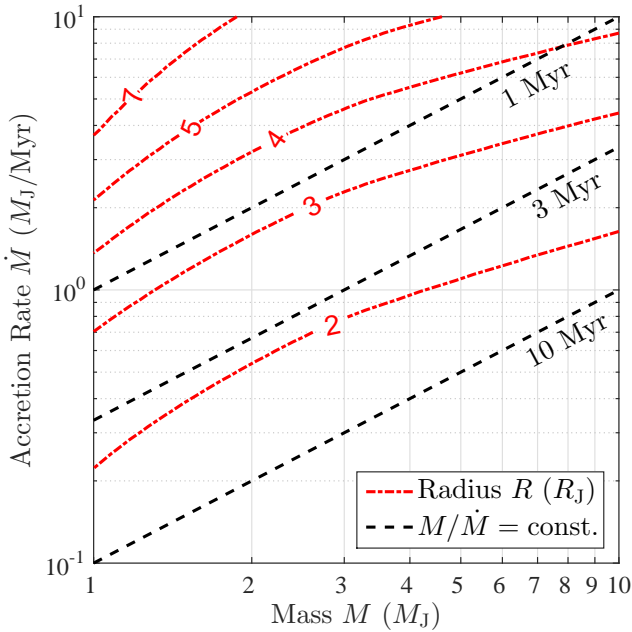}
	\caption{Same as Fig. \ref{fig:rad_contour}, but with a dust-free rcb opacity $\kappa=10^{-2}\textrm{ cm}^2\textrm{ g}^{-1}$. The constant-radius contours for this low-$\kappa$ model are dot--dashed and coloured red for consistency with the
	low-$\kappa$ curves in Fig. \ref{fig:lum_contour}.}
	\label{fig:radius_kappa}
\end{figure}

In Figs \ref{fig:rad_contour} and \ref{fig:radius_kappa} we 
exploit the same equality $t_{\rm cool}(M,R) = t_{\rm hydro} = M/\dot{M}$
to calculate $R$ for an arbitrary combination of $M$ and $\dot{M}$. 
That is,
we replace $t_{\rm hydro}$ with $M/\dot{M}$ in
equations \eqref{eq:tau_hydro}--\eqref{eq:tau_condition_hot} and solve for $R(M,\dot{M})$.
This approach dispenses with the need to prescribe
$\dot{M}(M)$ (i.e. $\beta$ in equation
\ref{eq:t_hydro} need not be specified).
The curves of constant
$R$ in Figs \ref{fig:rad_contour} and 
\ref{fig:radius_kappa} are nearly parallel
with the curves of constant accretion
time-scale $M/\dot{M}$, 
more so at long $M/\dot{M}$;
planets simply contract as they age 
(see also Figs \ref{fig:time} and \ref{fig:kappa_time}). 
For our high-opacity (dusty) model,
planets contract to $\sim$$7\,R_{\rm J}$ at
$M/\dot{M} = 3$ Myr
and to $\sim$$4\,R_{\rm J}$ at
$M/\dot{M} = 10$ Myr.
For our low-opacity (dust-free) model,
the canonical 1--2 $R_{\rm J}$ is 
attained for $M/\dot{M} \gtrsim 3$ Myr; 
the rate of contraction at longer times is overestimated because of our neglect of cold terms in the equation of state.

\begin{figure}
	\includegraphics[width=\columnwidth]{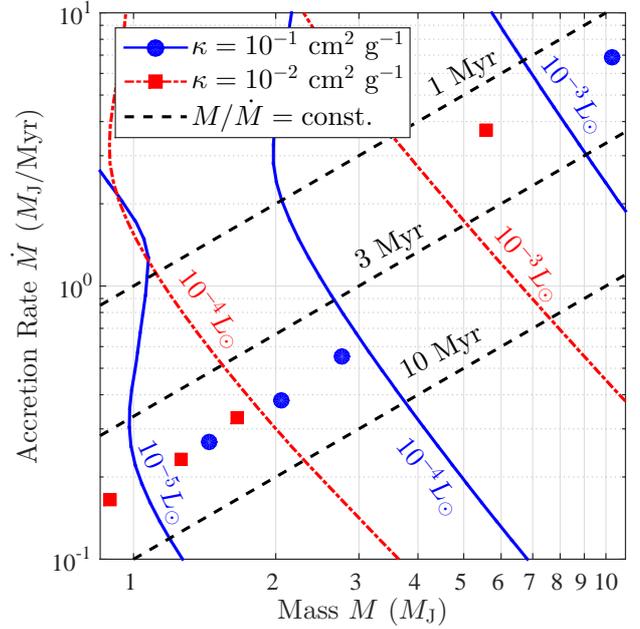} 
	\caption{Contours of constant accretion luminosity $L(M,\dot{M})=GM\dot{M}/R$ 
	computed for high $\kappa = 0.1\textrm{ cm}^2\textrm{ g}^{-1}$ (solid blue;
	the corresponding $R$ values are calculated in Fig. \ref{fig:rad_contour})
	and low $\kappa=10^{-2}\textrm{ cm}^2\textrm{ g}^{-1}$ (dot--dashed red; $R$ calculated in Fig. \ref{fig:radius_kappa}).
	Symbols denote four candidate planets, plotted using their measured luminosities $L$
	(so red squares are plotted in relation to red luminosity contours,
	and blue circles are in relation to blue contours)
	and ages $t$ that we equate to mass doubling times $M/\dot{M}$ (so plotted
	in relation to the dashed black lines). The plot maps 
	$(L,t=M/\dot{M})$ to $(M,\dot{M})$ and thus can be used to infer the
	masses and accretion rates of observed planets.
	From left to right, the planet
	candidates orbit PDS 70 (c and b; \citealt{Haffert2019,Keppler2018,Wagner2018}),
	HD 163296 \citep{Guidi2018}, and MWC 758 \citep{Reggiani2018}.
For PDS 70 we take the nominal age estimate from \citet{Muller2018} and the largest 
extinction assumed by \citet{Wagner2018}; weaker extinctions imply lower $L$ and therefore lower $M$. 
For PDS 70 b we assume the same conversion from H$\alpha$ luminosity to bolometric $L$ as in
\citet{Wagner2018}, and for PDS 70 c we scale from b using their K-band flux ratio,
observed to be equal to the H$\alpha$ ratio \citep{Haffert2019}.
For MWC 758 we assume the youngest age estimate; 
older ages lead to higher masses beyond the scope of our 
high-$\kappa$ model. In summary, the following values are used for the four objects in the plot (from left to right): $t/\textrm{Myr}=(5.4, 5.4, 5, 1.5)$ and $L/(10^{-5} L_\odot)=(2.1, 4.4, 8.2, 180)$.}
	\label{fig:lum_contour}
\end{figure}

In Fig. \ref{fig:lum_contour} we use the $R$ values from Figs \ref{fig:rad_contour} and \ref{fig:radius_kappa}
to calculate accretion luminosities $L$ as a function of $M$ and $\dot{M}$. A planet candidate's measured accretion luminosity falls
along a contour of constant $L$ (shown
in blue for high $\kappa$ and red for low $\kappa$),
constraining a combination of $M$ and $\dot{M}$---this combination is not merely the product $M \dot{M}$
because we do not assume a constant
$R$ but rather calculate its variation.
This joint constraint can be resolved
into separate individual constraints on $M$ and
$\dot{M}$ if we also measure the
system age and assume that it equals $M/\dot{M}$.
We perform this exercise
on four candidate planets:
PDS 70 b \citep{Keppler2018,Wagner2018},
PDS 70 c \citep{Haffert2019}, the infrared point source in HD 163296 \citep{Guidi2018}, and 
the infrared companion to MWC 758  \citep{Reggiani2018}.
These systems appear in Fig. \ref{fig:lum_contour} either as
blue circles or red squares, depending on whether
they are interpreted using our high-$\kappa$ or low-$\kappa$ model,
respectively. 
Since $R$ scales 
in our numerical model as $\kappa^{0.5}$ 
(cf. equation \ref{eq:kappa_r}), 
the inferred $M\dot{M}=RL/G$ varies
with $R$ and thus $\kappa$ at a fixed measured luminosity; hence the red points
must be paired with the red contours, and likewise for blue.

At face value, Fig. \ref{fig:lum_contour} indicates the planet candidates
around PDS 70 and HD 163296 are between $\sim$1--3 $M_{\rm J}$
and are accreting at rates of $\sim$0.2--0.6 $M_{\rm J}$/Myr;
in the case of MWC 758, $M\sim 5$--10 $M_{\rm J}$ and
$\dot{M} \sim 3$--7 $M_{\rm J}$/Myr.
There are, of course, numerous systematic uncertainties
underlying these estimates;
perhaps the most glaring is the poorly calibrated and
understood conversion of H$\alpha$ luminosity into bolometric accretion 
luminosity \citep{Rigliaco2012,Wagner2018}. We describe in the caption to
Fig. \ref{fig:lum_contour} the various assumptions we made
to assign ages and luminosities to our four objects.
It is nevertheless interesting and reassuring 
to infer from Figs
\ref{fig:rad_contour}--\ref{fig:lum_contour} 
a radius for PDS 70 b that lies between $1.8 R_{\rm J}<R<5
R_{\rm J}$, a range that overlaps with the radii fitted
to spectral energy distributions \citep{Muller2018}. In
particular, our high-$\kappa$ model yields a more slowly contracting
planet and can explain why some of the fitted
radii exceed $2 R_{\rm J}$, in contrast to low-$\kappa$ models 
\citep[e.g.][]{Mordasini2012}.
Our mass and radius ranges for PDS 70 b
are also consistent with the atmospheric `planet alone' models of \citet{ChristiaensAPJ} as fitted to the 
observed K-band spectrum; by comparison, their `planet + CPD' model yields a more massive best fit of $10\,M_{\rm J}$.
In addition, our inferred accretion rate agrees with the upper limit estimated from the non-detection
of Br$\gamma$ emission by \citet{ChristiaensMNRAS}. 

Earlier we saw that the high luminosity
of the companion to MWC 758 could be
explained by post-runaway accretion
in a viscous disk but not in an inviscid one.
The high inferred mass of MWC 758
gives another reason to disfavour its formation
in an inviscid disc. If $\beta=15$,
the planet mass cannot grow much beyond the
transition mass $M_0$, at most doubling
(Fig. \ref{fig:mass} and
fig. 3 of \citetalias{GC19} 
at 20 au).
For MWC 758, the transition mass of
$M_0 \approx 1 M_{\rm J}$ and the present-day
inferred mass of 5--10 $M_{\rm J}$ (Fig. \ref{fig:lum_contour}) are too far separated
to be reproduced by an inviscid evolution.

\section{Summary}\label{sec:summary}

Direct imaging is enabling the detection of planets
still accreting gas from their host circumstellar
discs. The luminosities of such planets are a key observable;
they constrain mass accretion rates and by extension planet formation theories.

What is essential to understand 
is how the planet's
gas accretion rate $\dot{M}$ drops from its runaway value, i.e.
how runaway stops 
while the planet is still embedded within its parent disc.
Previous treatments artificially limited the runaway $\dot{M}$ to some 
maximum value, and then lowered $\dot{M}$ to zero over an arbitrary
and unrealistically short time-scale, all to yield 
a prescribed final mass \citep{Marley2007, Mordasini2012}.
What we have posited instead is that gas accretion decays on a time-scale comparable to the lifetime of the gas disc. Post-runaway gas accretion may actually be the longest phase of gas giant formation.

What stops runaway? The opening of gaps is one possibility.
In the post-runaway era, gas accretion continues, 
but at a rate 
diminished by the reduction of disc
density by repulsive Lindblad torques. \citet{GC19}
show that gap opening causes the planet's mass doubling time-scale $M/\dot{M}$
to increase with increasing planet mass $M$, with the scaling
dependent on disc viscosity. In high-viscosity discs,
$M/\dot{M} \propto M^3$; in low-viscosity discs where gaps are deeper,
$M/\dot{M} \propto M^{15}$. We experimented here with both scalings
and found that they yielded qualitatively different luminosity evolutions. 
In the high-viscosity case, the accretion luminosity $L$ continues to rise
past runaway all the way to disc dispersal;
for
our nominal parameters, 
$L \sim 10^{-4}$--$10^{-3} L_\odot$ during the last few Myrs of the disc's
life. 
In the low-viscosity case, the luminosity drops post-runaway 
to $10^{-6}$--$10^{-5} L_\odot$. 
The divergence of luminosity tracks
is solely the result of different post-runaway 
accretion histories (which in turn stem from different
disc viscosities $\alpha$). The evolution during and prior to runaway
is otherwise identical between the two tracks,
as is the treatment of the accretion shock, which we do not take to be isothermal with the nebula, but which instead thermalizes at the blackbody temperature.
In this sense, our calculations are compatible with `hot start' models for post-formation cooling 
(see \citealt{Marleau2017} and \citealt{Marleau2019} for details).

The order-of-magnitude differences in $L$, which can be
understood analytically (see the appendix), persist for Myrs, and
highlight the importance of accurately modelling the terminal phase of
nebular accretion.
We established in Section \ref{sec:hydro} that during this
final phase, the planet contracts to a radius $R$ 
such that the Kelvin--Helmholtz cooling time $GM^2/(RL)$
is slaved to the (externally controlled) growth time $M/\dot{M}$.
From this equality of time-scales follows the planet's radius evolution with time
$R(t)$, and by extension the evolution of luminosity $L(t)=GM^2/(Rt)$ and 
photospheric temperature $T(t)=[L/(4\upi \sigma R^2)]^{1/4}$. 
Surface temperatures start from the ambient nebular temperature
during runaway, and increase post-runaway to values ranging
from $\sim$200--2000 K.

Even without specifying 
the full accretion history $\dot{M}(M)$,
we can solve for the
instantaneous $R$, $L$, and $T$
given the instantaneous $M$ and $\dot{M}$. The condition that
the planet's Kelvin--Helmholtz cooling time equals its
growth time, together with 
conditions at the radiative-convective boundary,
enables us to map $M$ and $\dot{M}$
to $R$ (Figs 
\ref{fig:rad_contour} and \ref{fig:radius_kappa}) 
and $L$ (Fig. \ref{fig:lum_contour}).
In the literature it is often assumed
that an accreting gas giant has $R \approx 2 R_{\rm J}$.
Our dust-free, low-opacity models can produce such
a radius, but our dusty, high-opacity models
yield radii that are generally larger by factors of a few.
Deciding between these models requires that we understand
how dust evolves in planetary atmospheres.

Finally, we applied our theory to the planet candidates
in PDS 70, HD 163296, and MWC 758.
From their observed luminosities $L$
and ages $t$ we inferred
masses $M \sim 1$--$10 \,M_{\rm J}$
and accretion rates $\dot{M} \sim 0.1$--$10 \,M_{\rm J}$/Myr
(Figure \ref{fig:lum_contour}).
These inferences rely on the assumption that
observed system ages can be interpreted as planet
mass doubling times: $t = M/\dot{M}$. 
We infer a radius for PDS 70 b between 1.8--$5 R_{\rm J}$ 
that is compatible with radius determinations from
spectral energy distributions \citep{Muller2018};
values greater than $2 R_{\rm J}$ are made possible
by higher opacities from dust 
which slow planetary contraction.

Future work needs to confront the angular momentum
barrier that spinning envelopes 
somehow
surmount in order to contract to their observed
sizes. The formation of rotationally
supported circumplanetary discs \citep[CPDs; e.g.][]{WardCanup2010} and associated
magnetic torques \citep{TakataStevenson96,Batygin2018}
are probably part of this story.
While CPDs yield
spectral energy distributions that would
differ in detail from those of the disc-less planets
that we have considered here
\citep[e.g.][]{Szulagyi2019},
bolometric luminosities
of CPDs and accreting planets should be comparable
(an accreting planet of radius $R$ has the same
bolometric luminosity as a disc accreting at the same
rate to an inner boundary layer of the same radius $R$).
Inferences of planet properties based on bolometric
luminosities, like the kind made in our
Fig. \ref{fig:lum_contour}, should be robust in this sense. 

\section*{Acknowledgements} 
We thank Ian Czekala, Ruobing Dong, Josh Eisner, Gabriel-Dominique Marleau, Mark Marley, Christoph Mordasini, Diana Powell, and Jason Wang for comments and discussions. We also thank the anonymous reviewer for suggestions which improved the paper. SG is supported by the Heising-Simons Foundation through a 51 Pegasi b Fellowship.




\bibliographystyle{mnras}
\input{lum.bbl}



\appendix
\section{Analytical scalings}\label{sec:analytical}

\subsection{Radiative envelope}\label{sec:anal_radiative}

During the early stages of contraction, the planet is still too cold for hydrogen molecules to dissociate, so $\gamma=7/5$.
In this case the optical depth at the rcb $\tau(M,R,T_0)$ is given by equation \eqref{eq:tau}, and equation \eqref{eq:tau_hydro} can be solved analytically to yield
\begin{equation}\label{eq:rad1}
\frac{\der\ln R}{\der\ln M}=-\frac{\beta(\gamma-1)+3-2\gamma}{4\gamma -5}=\begin{cases}
-7/3 & \beta=3\\
-31/3 & \beta=15\,,
\end{cases}
\end{equation}
with the normalization $R(M_0)=R_{\rm B}(M_0)\approx 2\times 10^3\,R_{\rm J}$. 
The luminosity is given by equation \eqref{eq:lum_radiative} and evolves as
\begin{equation}\label{eq:lum1}
\frac{\der\ln L}{\der\ln M}=-\frac{\beta(3\gamma-4)+7-6\gamma}{4\gamma-5}=\begin{cases}
+4/3 & \beta=3 \\
-8/3 & \beta=15\,.
\end{cases}
\end{equation}
The value of the luminosity at the end of runaway, $L(M_0)\approx 10^{-5} L_\odot$, with $L_\odot$ denoting the solar luminosity, is calculated by substituting $R(M_0)=R_{\rm B}(M_0)$ in equation \eqref{eq:lum_radiative}.

Interestingly, the luminosity can either decrease (if $\beta>\beta_{\rm cr}$) or increase (if $\beta<\beta_{\rm cr}$) during the planet's contraction, depending on the value of $\beta$. The critical value is
\begin{equation}\label{eq:beta_cr_cold}
\beta_{\rm cr}=\frac{6\gamma-7}{3\gamma-4}=7.
\end{equation}
The analytical solutions \eqref{eq:rad1}--\eqref{eq:lum1} fit well the earliest segments of the evolutionary tracks computed numerically with a more accurate $\gamma(\rho,T)$ (Fig. \ref{fig:mass}).

Using equations \eqref{eq:tau} and \eqref{eq:rad1}, the optical depth of the radiative envelope drops as
\begin{equation}\label{eq:tau_m}
\tau\propto\frac{1}{R}\left(\frac{R}{M}\right)^{1/(\gamma-1)}\propto
\begin{cases}
M^{-6} & \beta=3\\
M^{-18} & \beta=15\,
\end{cases}
\end{equation}
from an initial value of $\tau(M_0)=\kappa M_0/R_{\rm B}^2(M_0)\approx 4\times 10^2$.
When $\tau=1$ is reached, the 
diffusive radiative layer at $T\approx T_0$ vanishes and the planet becomes fully convective with a photospheric temperature $T>T_0$ (Section \ref{sec:convective}).

\subsection{Fully convective}\label{sec:anal_convective}

Most of the radiative-envelope phase can be calculated analytically with a single $\gamma=7/5$ as the planet remains fully molecular (Section \ref{sec:anal_radiative}). The transition to the fully-convective phase (filled circles in Figs \ref{fig:mass} and \ref{fig:time})
nearly coincides with the onset of molecular dissociation, so this phase can only be solved numerically (Section \ref{sec:convective}). 

We can still gain some analytical intuition by modelling the final stage of post-runaway accretion, when the planet's interior is fully ionized (after the two bumps in Figs \ref{fig:mass} and \ref{fig:time}, which indicate dissociation and ionization). During this phase, we model the planet piecewise:
$\gamma_1=5/3$ in the fully dissociated and ionized interior, and $\gamma_2=7/5$ in the molecular exterior where the photospheric temperature $T$ remains below the dissociation temperature $T_{\rm diss} \approx 2500$ K (see Fig. \ref{fig:mass}). The dissociation temperature depends only weakly on density according to the Saha equation. This two-piece construction reduces dissociation and ionization to a single sharp transition.

The integration of equation \eqref{eq:deriv} from $T<T_{\rm diss}$ to $T_c>T_{\rm diss}$ with the piecewise index yields
\begin{equation}\label{eq:piecewise}
\rho_{\rm rcb}\sim\frac{M}{R^3}\left(\frac{T}{T_{\rm diss}}\right)^{1/(\gamma_2-1)}\left(\frac{k_{\rm B}T_{\rm diss}R}{GM\mu}\right)^{1/(\gamma_1-1)}.   
\end{equation}
The optical depth scales as
\begin{equation}\label{eq:tau_hot}
\tau=\kappa\rho_{\rm rcb}h=\kappa\rho_{\rm rcb}\frac{k_{\rm B}TR^2}{GM\mu} \propto  T^{\gamma_2/(\gamma_2-1)}\frac{1}{R} \left(\frac{R}{M}\right)^{1/(\gamma_1-1)},
\end{equation}
where $T^4\propto M^{2-\beta}R^{-3}$ from equations \eqref{eq:t_hydro} and \eqref{eq:temp}.
By asserting that $\tau$ equals unity, equation \eqref{eq:tau_hot} reads
\begin{equation}\label{eq:rad_hot}
\frac{\der\ln R}{\der\ln M}=-\frac{\beta\gamma_2-2\gamma_2+4r}{7\gamma_2 -4-4r}=\begin{cases}
-19/17 & \beta=3\\
-103/17 & \beta=15\,,
\end{cases}
\end{equation}
where $r\equiv (\gamma_2-1)/(\gamma_1-1)=3/5$.
Using equations \eqref{eq:t_hydro} and \eqref{eq:temp}, the luminosity evolves as
\begin{equation}\label{eq:lum_hot}
\frac{\der\ln L}{\der\ln M}=\frac{2(16-5\beta)}{17}=
\begin{cases}
+2/17 & \beta=3 \\
-118/17 & \beta=15\,,
\end{cases}
\end{equation}
with the critical $\beta_{\rm cr}=16/5$ during this stage.

We reiterate that equations \eqref{eq:rad_hot} and \eqref{eq:lum_hot} are applicable only after the planet's interior becomes fully ionized. Even then, these scaling relations are approximate, as they do not account for the (weak) dependence of $T_{\rm diss}$ on density, and treat dissociation and ionization as a single sharp transition.
Nevertheless, the equations reproduce the qualitative behaviour of the radius and luminosity close to the time of disc dispersal (filled squares in Fig. \ref{fig:mass}).
Moreover, equations \eqref{eq:temp} and \eqref{eq:tau_hot} specify the dependence of the planet's radius, at a given time, on the opacity:
\begin{equation}\label{eq:kappa_r}
\frac{\der\ln R}{\der\ln \kappa}=\left(\frac{3}{4}\frac{\gamma_2}{\gamma_2-1}-\frac{1}{\gamma_1-1}+1\right)^{-1}=\frac{8}{17}\,,   
\end{equation}
which helps explain the results of Section \ref{sec:opacity}.

Equations \eqref{eq:rad1}--\eqref{eq:lum1} and \eqref{eq:rad_hot}--\eqref{eq:lum_hot} illustrate the sensitivity of our results to $M_0$, the transition mass from runaway to post-runaway. Since the slopes given by these equations depend only on $\beta$ and $\gamma$, the curves for different $M_0$ are, to leading order, parallel, departing from different points along the dot--dashed red line in Fig. \ref{fig:mass}. 


\bsp	
\label{lastpage}
\end{document}